\documentclass{article}
\usepackage{cite}
\usepackage{amsmath,amssymb,amsfonts}
\usepackage{algorithmic}
\usepackage{graphicx}
\usepackage{textcomp}
\usepackage{xcolor}
\usepackage{textgreek}
\usepackage{mathtools}
\usepackage{engord}
\usepackage{url}
\usepackage{enumitem}
\usepackage{calc}
\usepackage{subfig}

\def\BibTeX{{\rm B\kern-.05em{\sc i\kern-.025em b}\kern-.08em
    T\kern-.1667em\lower.7ex\hbox{E}\kern-.125emX}}

\providecommand{\keywords}[1]
{
  \small	
  \textbf{\textit{Index Terms---}} #1
}

\begin{document}

\title{Deterministic Random Number Generator Attack against the Kirchhoff-Law-Johnson-Noise Secure Key Exchange Protocol\\}

\author{Christiana Chamon\textsuperscript{1}, Shahriar Ferdous\textsuperscript{2}, and Laszlo Kish\textsuperscript{3}\\
\textit{Department of Electrical and Computer Engineering} \\
\textit{Texas A\&M University}\\
College Station, TX 77843-3128, USA \\
cschamon@tamu.edu, ferdous.shahriar@tamu.edu, laszlokish@tamu.edu\\
ORCiD: \textsuperscript{1}0000-0003-3366-8894, \textsuperscript{2}0000-0001-5960-822X, \textsuperscript{3}0000-0002-8917-954X}

\date{}

\maketitle

\begin{abstract}
This paper demonstrates the vulnerability of the Kirchhoff-Law-Johnson-Noise (KLJN) secure key exchanger to compromised random number generator(s) even if these random numbers are used solely to generate the noises emulating the Johnson noise of Alice's and Bob's resistors. The attacks shown are deterministic in the sense that Eve's knowledge of Alice's and/or Bob's random numbers is basically deterministic. Moreover, no statistical evaluation is needed, except for rarely occurring events of negligible, random waiting time and verification time. We explore two situations. In the first case, Eve knows both Alice's and Bob's random noises. We show that, in this situation, Eve can quickly crack the secure key bit by using Ohm’s Law. In the other situation, Eve knows only Bob’s random noise. Then Eve first can learn Bob's resistance value by using Ohm’s Law. Therefore, she will have the same knowledge as Bob, thus at the end of the bit exchange period, she will know Alice's bit.
\end{abstract}


\section{Introduction}
\label{intro}
This paper will introduce the concept of secure communications, the KLJN scheme, and random number generators, and present the new theoretical attacks on the KLJN scheme based on compromised random number generators.

\subsection{On Secure Communications}
\begin{figure}[htbp]
\centerline{\includegraphics[width=8.25cm]{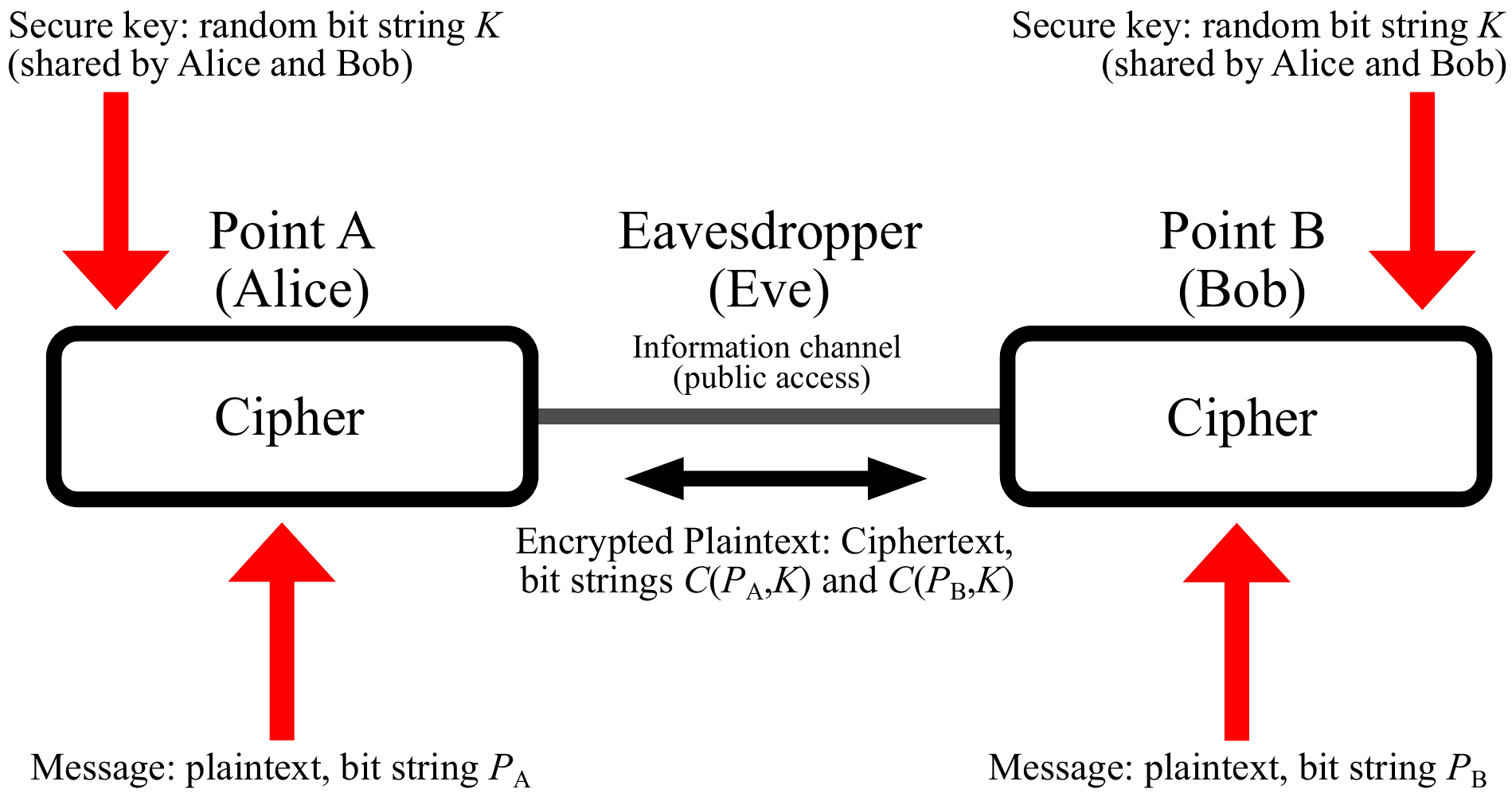}}
\caption{Symmetric-key cryptography \cite{b1}. Alice and Bob use ciphers to exchange secure keys, or a string of bits, through a public channel. The ciphers encrypt plaintext, or convert it into ciphertext. The secure key is represented by $K$, the plaintext messages of Alice and Bob are represented by $P_\mathrm{A}$ and $P_\mathrm{B}$, respectively, and the ciphertext is represented as a function of $P$ and $K$, or $C$($P$,$K$).}
\label{symmetric}
\end{figure}

One way to establish the security of a communication is through encryption, that is, the conversion of plaintext into ciphertext via a cipher \cite{b1}. Fig.~\ref{symmetric} provides the general scope of symmetric-key cryptography \cite{b1}. The key is a string of random bits and both communicating parties Alice and Bob use the same key and ciphers to encrypt and decrypt their plaintext.

For a plaintext message $P$ and a secure key $K$, the encrypted message, or the ciphertext $C$, is a function of $P$ and $K$, that is,

\begin{equation}
C=C(P,K).\label{ciphertext}
\end{equation}

In symmetric-key cryptography, for decryption, the inverse operation is used:

\begin{equation}
P=C^{-1}[C(P,K),K].\label{plaintext}
\end{equation}

\noindent Because the secure keys must be the same at the two sides (shared secret), another type of secure data exchange is needed before the encryption can begin: the secure key exchange, which is the generation and distribution of the secure key over the communication channel. Usually, this is the most demanding process in the secure communication because the communication channel is accessible by Eve thus the secure key exchange is itself a secure communication where the cipher scheme shown in Fig.~\ref{symmetric} cannot be used. Eve records the whole communication during the key exchange, too. She knows every detail of the devices, protocols, and algorithms in the permanent communication system (as stated by Kerckhoffs's principle/Shannon’s maxim \cite{b2}), except for the key. In the ideal case of perfect security, the key is securely generated/shared, immediately used by a One Time Pad \cite{b3}, and discarded after the usage. In practical cases, usually there are deviations from these strict conditions, yet the general rule holds: a secure system cannot be more secure than its key.

The key is assumed to be generated from truly random numbers. Any predictability of the key leads to compromised security \cite{b3}. In this paper, we demonstrate attacks on the unconditionally secure Kirchhoff-Law-Johnson-Noise (KLJN) symmetric-key exchange based on compromised random number generators (RNGs).

\subsection{The KLJN Scheme}
\label{KLJN}
The KLJN scheme \cite{b41,b42,b43,b44,b45,b46,b47,bVMG,b48,b49,b50,b51,b52,b53,b54,b55,b56,b57,b58,b59,b60,b61,b62,b63,b64,b65,b66,b67,b68,b69,b70,b71,b72,b73,b74,b75,b76,bMutaz4,b77,b78,b79,b80,b81,b82,b83,b84,b85,b86,b87,b88,b89,b90,b91,b92} is a statistical physical scheme based on the thermal noise of resistors. It is a classical (statistical) physical alternative of Quantum Key Distribution (QKD). Note, in papers  \cite{b3,b4,b5,b6,b7,b8,b9,b10,b11,b12,b13,b14,b15,b16,b17,b18,b19,b20,b21,b22,b23,b24,b25,b26,b27,b28,b29,b30,b31,b32,b33,b34,b35,b36,b37,b38,b39,b40}, important criticisms and attacks are presented about QKD indicating some of the most important difficulties of unconditionally secure quantum hardware and their theory.

\begin{figure}[htbp]
\centerline{\includegraphics[width=8.25cm]{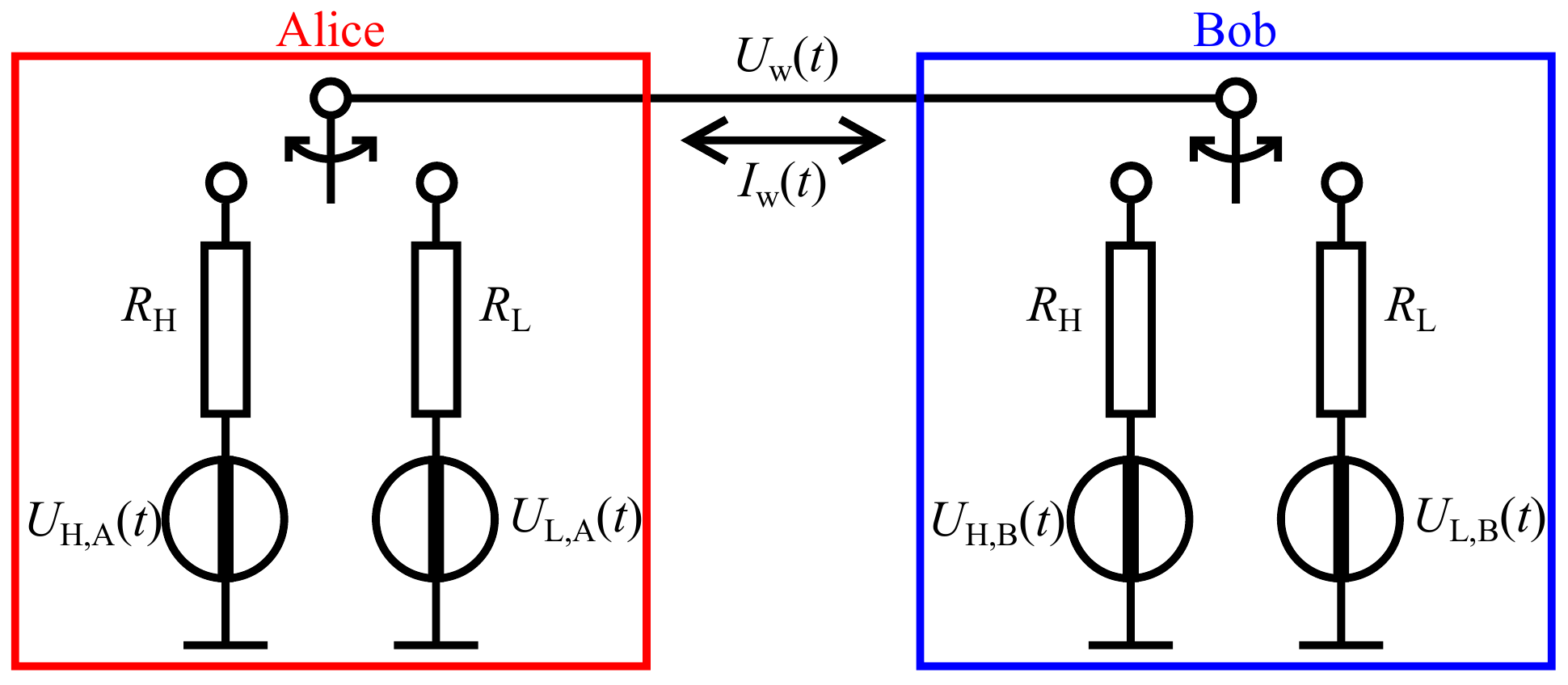}}
\caption{The core of the KLJN scheme. The two communicating parties, Alice and Bob, are connected via a wire. The wire voltage and current are denoted as $U_\mathrm{w}(t)$ and $I_\mathrm{w}(t)$, respectively. The parties have identical pairs of resistors $R_\mathrm{H}$ and $R_\mathrm{L}$ ($R_\mathrm{H} > R_\mathrm{L}$) that are randomly selected and connected to the wire at the beginning of the bit exchange period. The statistically independent thermal noise voltages $U_\mathrm{H,A}(t)$, $U_\mathrm{L,A}(t)$, and $U_\mathrm{H,B}(t)$, $U_\mathrm{L,B}(t)$ represent the noise voltages of the resistors $R_\mathrm{H}$ and $R_\mathrm{L}$ of Alice and Bob, respectively.}
\label{figKLJN}
\end{figure}

Fig.~\ref{figKLJN} illustrates the core of the KLJN scheme. The two communicating parties, Alice and Bob, are connected via a wire. They have identical pairs of resistors, $R_\mathrm{A}$ and $R_\mathrm{B}$. The statistically independent thermal noise voltages $U_\mathrm{H,A}(t)$, $U_\mathrm{L,A}(t)$, and $U_\mathrm{H,B}(t)$, $U_\mathrm{L,B}(t)$ represent the noise voltages of the resistors $R_\mathrm{H}$ and $R_\mathrm{L}$ ($R_\mathrm{H} > R_\mathrm{L}$) of Alice and Bob, respectively, which are generated from random number generators (RNGs) and must have a Gaussian distribution \cite{b54,b57}.

At the beginning of each bit exchange period (BEP), Alice and Bob randomly choose one of their resistors to connect to the wire. The wire voltage $U_\mathrm{w}(t)$ and current $I_\mathrm{w}(t)$ are as follows:

\begin{equation}
U_\mathrm{w}(t)=I_\mathrm{w}(t)R_\mathrm{B}+U_\mathrm{B}(t),\label{Uw}
\end{equation}

\begin{equation}
I_\mathrm{w}(t)=\frac{U_\mathrm{A}(t)-U_\mathrm{B}(t)}{R_\mathrm{A}+R_\mathrm{B}}\label{Iw}
\end{equation}

\noindent where $U_\mathrm{A}(t)$ and $U_\mathrm{B}(t)$ denote the instantaneous noise voltage of the resistor chosen by Alice and Bob, respectively. Alice and Bob (as well as Eve) use the mean-square voltage of the wire to assess the bit status, given by the Johnson formula

\begin{equation}
U_{\mathrm{w,eff}}^2=4kT_{\mathrm{eff}}R_{\mathrm{p}}\Delta{f_{\mathrm{B}}}\label{meansquare}
\end{equation}

\noindent where $k$ is the Boltzmann constant (1.38 x 10\textsuperscript{-23}~J/K), $T_\mathrm{eff}$ is the publicly agreed effective temperature, $R_\mathrm{P}$ is the parallel combination of Alice and Bob’s chosen resistors, given by

\begin{equation}
R_{\mathrm{P}}=\frac{R_{\mathrm{A}}R_\mathrm{B}}{R_\mathrm{A}+R_\mathrm{B}},\label{Rp}
\end{equation}

\noindent and $\Delta$$f_\mathrm{B}$ is the noise bandwidth of the generators.

\begin{figure}[htbp]
\centerline{\includegraphics[width=8.25cm]{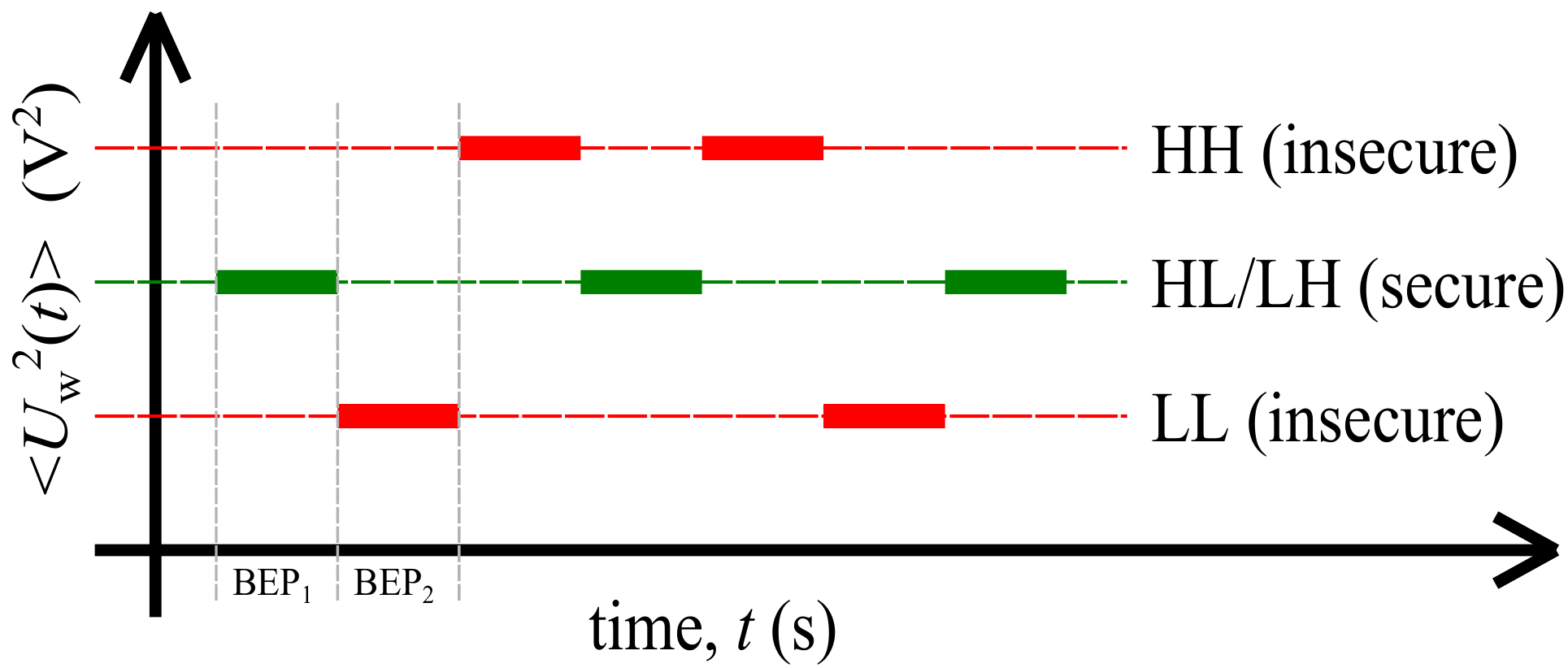}}
\caption{The three mean-square voltage levels. The HH and LL cases represent insecure situations because they form a unique mean-square voltage. The HL and LH cases represent secure bit exchange because Eve cannot distinguish between the two corresponding resistance situations while Alice and Bob can because they know their own connected resistance values.}
\label{threelevels}
\end{figure}

Four possible resistance situations can be formed by Alice and Bob: HH, LL, LH, and HL. Using the Johnson formula, these correspond to three mean-square voltage levels, as shown in Fig.~\ref{threelevels}. 

The HH and LL cases represent insecure situations because they form a unique mean-square voltage. These insecure exchange values are discarded by Alice and Bob. The HL and LH cases represent secure bit exchange because Eve cannot distinguish between the two corresponding resistance situations (LH or HL), while Alice and Bob can.

Several attacks against the KLJN system have been proposed \cite{b73,b74,b75,b76,bMutaz4,b77,b78,b79,b80,b81,b82,b83,b84,b85,b86,b87,b88,b89,b90,b91,b92}, but no attack has been able to compromise its information-theoretic (unconditional) security because each known attack is either invalid (conceptually incorrect errors in theory and/or experiments) or can be nullified by a corresponding defense scheme such that Eve’s information entropy about the key approaches the bit length of the key, while Alice’s and Bob’s information entropy about the key approaches zero. None of the attacks against the KLJN scheme are RNG attacks, thus the new attacks presented in this paper are based on the assumption that the random number generators Alice and Bob use to generate their noises are compromised.

\subsection{Random Number Generator Attacks}
\label{RNG}
There are two classes of practical random number generators: true (physical) and computational. The nature of computational RNGs is that they collect randomness from various low-entropy input streams and try to generate outputs that are in practice indistinguishable from truly random streams \cite{b93,b94,b95,b96,b97,b98}. The randomness of an RNG relies on the uncertainty of the random seed, or initialization vector, and a long sequence with uniform distribution. The moment an adversary learns the seed, the outputs are known, and the RNG is compromised.

Various RNG attacks exist against conditionally secure communications \cite{b93,b94,b95,b96,b97,b98}. Unconditionally secure communications also require true random numbers for perfect security, and that is also true for the noises of Alice and Bob and for the randomness of their switch-driving, which uses a different RNG from the voltage sources. However, it is unclear how Eve can utilize compromised RNGs, from which she has obtained by spying on the root of the RNGs, to attack the KLJN scheme. The RNG outputs must imitate thermal noise, and in many practical applications, only a computational RNG is possible. Here, we demonstrate with two simple attack examples that compromised noises lead to information leak.

The rest of this paper is organized as follows. Section~\ref{experiment} describes two new attack protocols and the waiting/verification (no-response) time scenario, and Section~\ref{conclusion} concludes this paper.

\section{Attack Methodology}
\label{experiment}

Two theoretical situations are introduced where Eve can use compromised RNGs to crack the KLJN scheme: one where Eve knows the roots of both Alice’s and Bob’s generators (bilateral parameter knowledge), and another where Eve knows the root of only Bob's generator (unilateral parameter knowledge). A scenario is also introduced where the only statistical evaluation needed is in the rarely-occurring event of a negligible waiting/verification (no-response) time.

\subsection{Bilateral Parameter Knowledge}
\label{bilateral}
Eve knows the roots of both Alice’s and Bob’s RNGs, thus she knows the instantaneous amplitudes of the noise voltage generators for each of their resistors (see Fig.~\ref{figKLJN}). With $U_{\mathrm{H,A}}(t)$, $U_{\mathrm{L,A}}(t)$, $U_{\mathrm{H,B}}(t)$, and $U_{\mathrm{L,B}}(t)$ known, Eve measures the wire voltage $U_{\mathrm{w}}(t)$ and $I_{\mathrm{w}}(t)$ and the wire current $I_{\mathrm{w}}(t)$ and $I_{\mathrm{w}}(t)$ to determine $R_{\mathrm{B}}$ (see \eqref{Uw}) and determines that the correct waveform for $R_{\mathrm{B}}$ is a flat line at the expected resistor value, while the incorrect waveform is a noise with divergent spikes.

\begin{figure}[htbp]
\centerline{\includegraphics[width=8.25cm]{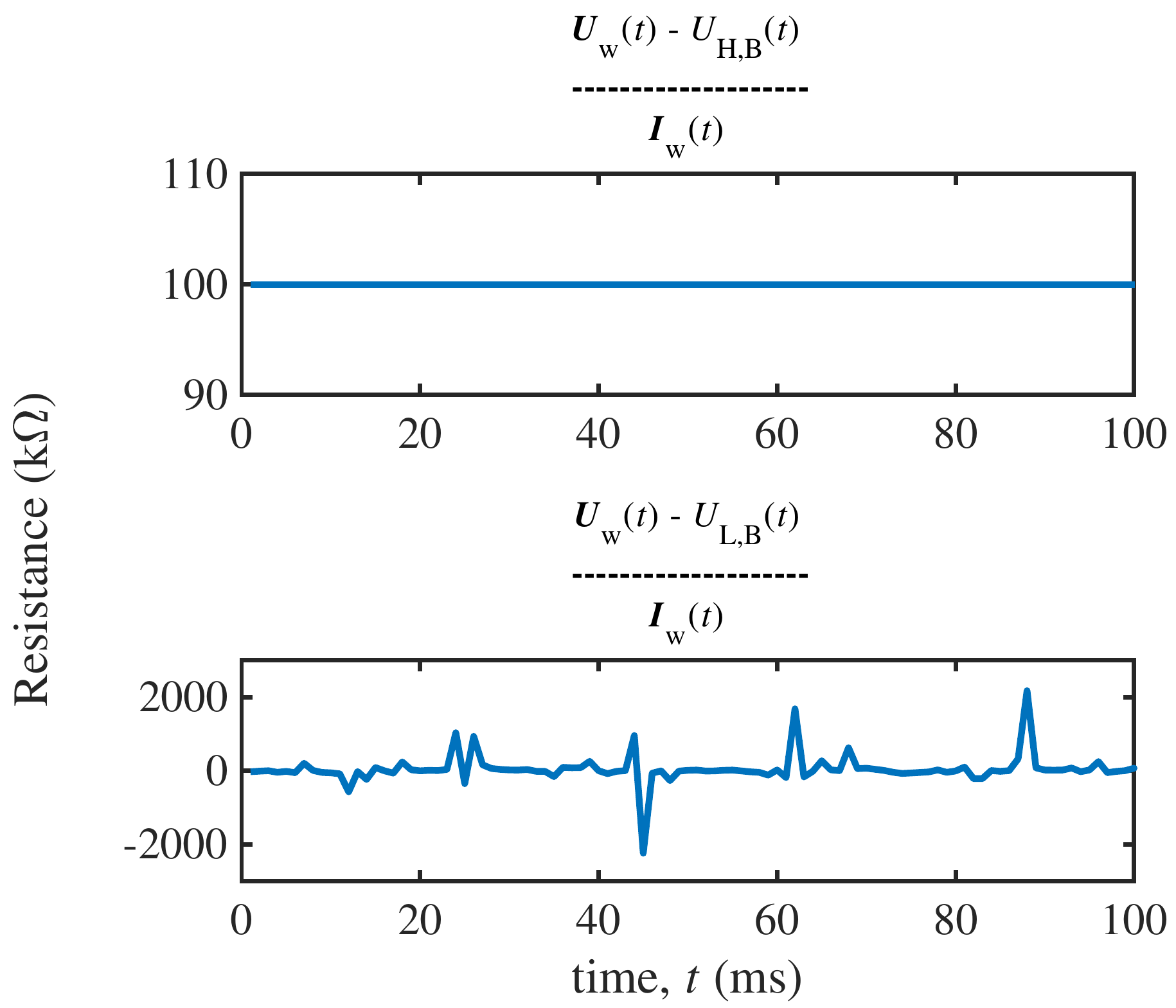}}
\caption{\ A realization of the waveforms for $R_{\mathrm{B}}$ (see \eqref{Uw}) at $R_{\mathrm{H}} = 100$~k\textOmega, $R_{\mathrm{L}} = 10$~k\textOmega, $T_{\mathrm{eff}} = 10^{18}$~K, and $\Delta f_{\mathrm{B}} = 500$~Hz. The flat line corresponds to $R_{\mathrm{H}}$, thus Eve determines that Bob has chosen $R_{\mathrm{H}}$. The incorrect waveform is a noise with divergent spikes.}
\label{RB}
\end{figure}

A realization of the waveforms for $R_{\mathrm{B}}$ is shown in Fig.~\ref{RB} at $R_{\mathrm{H}} = 100$~k\textOmega, $R_{\mathrm{L}} = 10$~k\textOmega, $T_{\mathrm{eff}} = 10^{18}$~K, and $\Delta f_{\mathrm{B}} = 500$~Hz. At the present situation, the flat line corresponds to $R_{\mathrm{H}}$, thus Eve determines that Bob has chosen $R_{\mathrm{H}}$. With $U_{\mathrm{A}}(t)$, $U_{\mathrm{B}}(t)$, $R_{\mathrm{B}}$, and $I_{\mathrm{w}}(t)$ known, Eve uses \eqref{Iw} to solve for $R_{\mathrm{A}}$.

\subsection{Unilateral Parameter Knowledge}
\label{unilateral}
Eve knows only the root of Bob's generator RNG, thus she knows merely the noise generator outputs of Bob's resistors, $U_{\mathrm{L,B}}(t)$ and $U_{\mathrm{H,B}}(t)$. Alice's generator voltages are unknown to her. Eve uses the same protocol as mentioned in the bilateral case (see Section~\ref{bilateral}) to determine $R_{\mathrm{B}}$, but because Alice’s generator voltages are unknown to her, she cannot use \eqref{Iw} to solve for $R_{\mathrm{A}}$. Instead, she uses the KLJN protocol as Bob does: use the entire bit exchange period to evaluate the measured mean-squared voltage on the wire. From that value, by using \eqref{meansquare}, she evaluates the parallel resultant $R_{\mathrm{P}}$ of the resistances of Alice and Bob. From $R_{\mathrm{P}}$ and $R_{\mathrm{B}}$, she can calculate $R_{\mathrm{A}}$.

\subsection{Timing}
\label{timing}
In each attack, Eve may or may not know which noise belongs to which resistor, or the noises may provide a voltage-to-current ratio that results in either $R_{\mathrm{H}}$ or $R_{\mathrm{L}}$. In these scenarios, there will be a negligible waiting/verification (no-response) time before Eve can differentiate between the noises. The probability of the two independent noises being equal at any given point is

\begin{equation}
p_{\mathrm{0}}=\frac{1}{2^\Delta},\label{p0}
\end{equation}

\noindent where $\Delta$ is the number of resolution bits.

We can consider the event that the noises run identically for $n$ subsequent steps to be a geometric distribution, with a probability of

\begin{equation}
p=p_0^n,\label{p1}
\end{equation}

\noindent where $n=t/\tau$ and $\tau$ is the autocorrelation time, given by the Nyquist Sampling Theorem

\begin{equation}
\tau\approx\frac{1}{2\Delta f_{\mathrm{B}}}.\label{tau}
\end{equation}

\noindent The approximation sign is due to the quantized nature of the noises, as they are constant throughout the bit exchange periods.

While such an identical match between the two noises is taking place, Eve cannot distinguish between the two noises, thus the actual resistance situation (see Fig.~\ref{threelevels}) remains secure. However, the exponential decay in \eqref{p1} yields an efficient cracking scheme of the secure key bit value within a short amount of time. In accordance with \eqref{p1}, the probability of the two independent seeds running identically is

\begin{equation}
p\approx p_0^{t/\tau}\approx\left(\frac{1}{2^\Delta}\right)^{t/\tau}.\label{p}
\end{equation}

\section{Conclusion}
\label{conclusion}
Secure key exchange protocols utilize random numbers, and compromised random numbers lead to information leak. So far, it had been unknown how Eve can utilize compromised random number generators to attack the KLJN protocol. To demonstrate how compromised RNGs can be utilized by Eve, we have introduced two simple attacks on the KLJN scheme.

We showed that if Eve knows the root of both Alice’s and Bob’s RNGs, that is, when she exactly knows the random numbers, she can use Ohm’s Law to crack the bit exchange. Eve can extract the bit very quickly, and she will learn the exchanged bit faster than Alice and Bob who know only their own random numbers.

We have also shown that if Eve knows the seed of only Bob’s RNG, she can still use Ohm’s Law to crack the secure bit. However, she is required to utilize the whole bit exchange period.

No statistical evaluation is needed, except for in the rarely-occurring event that Eve does not know which RNG belongs to which resistor, which will render in a waiting/verification (no-response) time that has a negligible effect on Eve’s cracking scheme.

It is important to note that:
\begin{itemize}
\item To utilize these attacks, we implicitly used Kerckhoffs’s principle \cite{b2}, which means Eve knows all the fine details of the protocol, including how the seeds are utilized and the RNG outputs timed.
\item This demonstration was done assuming an ideal KLJN scheme. Future work would involve a practical implementation with a cable simulator and related delays and transients.
\item A deterministic knowledge of the random number(s) by Eve is a strong security vulnerability. However, it is an illustrative way how such attacks can be developed.
\item Open problems are new attack schemes where Eve's knowledge of the RNGs is only statistical \cite{b100}.
\end{itemize}

\end{document}